\documentclass[12pt,twoside]{article}

\usepackage{graphicx}

\begin{document}

\title{SEMI-ANALYTICAL ANALYSIS OF\\
HELIUM SYNTHESIS IN BRANE COSMOLOGY}

\author{J.C. Fabris\thanks{%
e-mail: fabris@cce.ufes.br} and J.A.O. Marinho\thanks{%
e-mail: admarinho@hotmail.com}\\
Departamento de F\'{\i}sica, Universidade Federal do Esp\'{\i}rito Santo, \\
CEP29060-900, Vit\'oria, Esp\'{\i}rito Santo, Brazil\\} \maketitle

\begin{abstract}
The problem of primordial nucleosynthesis of helium in brane
cosmology is studied using a semi-analytical method, where the
Fermi-Dirac statistic is ignored. This semi-analytical method
agrees with a more complete numerical calculation with a precision
of order of $10\%$ or better. The quadratic term for the matter
density is the only source considered in the modified Einstein
equations predicted by the brane configuration. This hypothesis is
justified a posteriori. An agreement between theoretical and
observational values for the helium abundance is obtained if the
fundamental mass scale in five dimensions is of the order of $M
\sim 5\times10^3\,GeV$.

\vspace{0.7cm}


\end{abstract}

\section{Introduction}

The brane model is one of the most interesting cosmological
scenario proposed in recent times \cite{dvali1,dvali2,dvali3}. For
a review of brane cosmology, see references \cite{langlois,brax}.
In its most simple formulation, a five dimensional space-time is
considered where a three-brane configuration is settled out. The
ordinary matter and fields are restricted to "live" on the brane.
Only gravity and a cosmological constant term are allowed to
"live" in the entire four dimensional space, called the {\it
bulk}. One fundamental aspect of this construction is the fact
that the fifth dimension may be compact but not small: This allows
to solve the hierarchy problem, explaining why the Planck mass in
our usual four dimensional space-time takes the huge value $M_{Pl}
\sim 10^{19}\,GeV$. In fact, the Planck mass $M_{Pl}$ is related
with the five dimensional fundamental mass scale $M$ through the
relation
\begin{equation}
M^2_{Pl} = M^{3}R \quad ,
\end{equation}
where $R$ is the characteristic dimension of the fifth dimension.
Hence, the bulk gravitational coupling can have values compatible
with other physical coupling constants provided the scale $R$ is
very large, inducing at same time a huge value for the Planck mass
on the brane. In a very fashion variation of this model, Randall
and Sundrum \cite{randall1,randall2} had considered a five
dimensional anti-deSitter model (a negative cosmological constant
living in the bulk), where gravity has the usual features on the
brane even if the fifth dimension is infinite.
\par
One of the distinguishing features of brane cosmology is the
presence of a quadratic matter term as source of the Einstein
equations which describes the gravitational dynamics on the brane.
This quadratic term comes from the matching conditions of the five
dimensional Einstein equations in presence of a brane, together
with the assumption that ordinary matter is restricted to "live"
on the brane. Besides this quadratic term, there is a dark
radiation term (which may contribute with negative values for the
source terms) as well as a cosmological constant, if a
cosmological term is allowed in the bulk. Hence, the dynamics for
our Universe is very different from that predicted by the
traditional standard cosmological model. Due to the quadratic
character of the new ordinary source term, its effect is in
principle important during the very early Universe: The
contribution of the quadratic term decreases very rapidly as the
Universe expands.
\par
Many cosmological consequences may be extracted from this
primordial scenario, and the brane cosmology became the object of
intensive studies in these last times. Here, we will address one
important point concerning the brane scenario: The primordial
nucleosynthesis. In fact, the presence of a quadratic matter term
and of a dark radiation term may have many consequences for the
abundance of the primordial light elements, like deuterium and
helium. Such problem has already been treated in references
\cite{ichiki,bratt}. However, in reference \cite{ichiki} mainly
the question of the influence of dark energy has been investigate;
in reference \cite{bratt}, on the other hand, a balance between
the contribution of the quadratic term and dark energy is required
in order that the traditional scenario for nucleosynthesis is not
substantially affected. These authors have kept the main features
of primordial nucleosynthesis process, like the energy scales. In
reference \cite{bratt} it has been concluded that large negative
values for the dark radiation are allowed.
\par
In the present work we will take two main different
considerations. First, we will emphasize the role of the quadratic
term in the source of the Einstein equations. This quadratic term
changes drastically the temperature curve, and the characteristic
energy scales under which the nucleosynthesis takes place becomes
completely different with respect to the standard model. In
particular, it will be shown that the observed value for the
helium abundance implies that the nucleosynthesis is effective
between $100\,MeV < T < 0.1\,MeV$, in contrast with the usual
values $1\,MeV < T < 0.1\,MeV$. The fundamental mass scale which
leads to values for the helium abundance compatible with
observational data is of the order of $M \sim 5\times10^{3}\,GeV$.
In this case, the dark radiation and cosmological constant may
indeed be neglected at least in what concerns nucleosynthesis. In
some sense, the analysis performed here is more restrictive than
those made in references \cite{ichiki,bratt} since these authors
have determined a lower bound for the fundamental mass scale.
However, we must stress that a more general analysis taking into
account the contribution of all terms in the brane model, which
asks for a substantial modification of the semi-analytical method,
may lead to higher values for the fundamental mass scale.
\par
The other aspect that distinguish the present work from the
previous ones concerns the employment of the semi-analytical
analysis for the helium synthesis which has been developed in
reference \cite{bernstein}. This semi-analytical analysis, which
avoids the use of large numerical codes, is possible if the
Fermi-Dirac statistic is ignored. Even if such statistics seems to
be very relevant in the evaluation of the transmutation process
that occurs in the primordial nucleosynthesis, it does not affect
sensibly the final results: The semi-analytical method leads to a
value of the helium abundance only some per cents different from
that obtained when the Fermi-Dirac statistic is taken into
account. A very important advantage of using such semi-analytical
method concerns the fact that the main physical steps in the
computation of the helium abundance may be emphasized.
\par
This article is organized as follows. In the next section, we
sketch the main features of brane cosmology. In section 3, the
semi-analytical method is exposed in its general lines. Its
application to the brane cosmology is presented in section 4,
while in section 5 the conclusions are presented.

\section{Brane cosmology}

The brane model to be considered here is the simplest one, which
keeps the main features of more complicated models. This model is
constructed in a five-dimensional space-time, where the fifth
coordinate is not compact. This marks a strinking difference with
respect to the usual Kaluza-Klein models. In this model, our usual
world is constrained to a three-brane, where ordinary matter and
fields "live". Gravity, on the other side, is defined in the
entire five-dimensional space-time. The field equations read
\begin{equation}
G_{AB} = \kappa T_{AB} + \Lambda g_{AB} \quad ,
\end{equation}
where $A,B = 0,1,...,4$, $\Lambda$ is the cosmological constant which also "lives"
in the whole space-time and $\kappa$ is the gravitational coupling in five-dimensions.
The five-dimensional space-time is called {\it bulk}.
The metric describing this configuration is
\begin{equation}
ds_5^2 = n^2(t,y)dt^2 - a^2(t,y)\gamma_{ij}dx^idy^j - dy^2 \quad .
\end{equation}
The energy-momentum tensor takes the form
\begin{equation}
T^A_B = diag(\rho_b, - p_b, - p_b, - p_b, 0)\delta(y)
\end{equation}
where the delta function assures that ordinary matter is confined
on the brane which is located at $y = 0$.
\par
The presence of the brane implies the need to impose match
conditions through the extrinsic curvature. Using the field
equations and considering these match conditions, we end up with
the following equation driving the dynamics of the scale factor
and matter on the brane:
\begin{equation}
\label{em1}
\biggr(\frac{\dot a_0}{a_0}\biggl)^2 = \frac{\kappa^4}{36}\rho_b^2 + \frac{\Lambda}{6}
- \frac{k}{a_0^2} + \frac{{\cal C}}{a_0^4} \quad ,
\end{equation}
the subscript $0$ indicating that all the quantities are evaluated on the brane.
This subscript will be ignored henceforth. As usual,
$k$ denotes the curvature of the spatial section, and ${\cal C}$ is an integration constant
generally called {\it dark radiation}, since it appears in the equation of motion similarly to
a radiation term. Notice, however, that ${\cal C}$ may take any sign. An important
characteristic of brane cosmology is the fact that the matter term appears quadratically in
the equation of motion.
In what follows, we neglect the curvature, fixing $k = 0$.
\par
The standard Friedmann-Robertson-Walker cosmology is characterized
by a linear matter term, in opposition to equation (\ref{em1}).
However, it is possible in brane cosmology to generate such linear
term. This can be achieved by redefining the matter term as
$\rho_b = \rho + \sigma$, where $\sigma$ is a constant. Hence,
(\ref{em1}) becomes
\begin{equation}
\label{em2}
\biggr(\frac{\dot a}{a}\biggl)^2 = \frac{\kappa^4}{36}\rho^2  + \frac{\kappa^4\sigma}{18}\rho
+ \frac{\kappa^4\sigma^2}{36} + \frac{\Lambda}{6}
+ \frac{{\cal C}}{a^4} \quad .
\end{equation}
If we make the identification $8\pi G = \kappa^4\sigma/18$, the ordinary matter term is
reproduced besides the quadratic one. At same time, a new cosmological term appears.
The total cosmological term may be set equal zero by fine tuning their values. This
is possible mainly in the context of the Randall-Sundrum model, where an anti-deSitter
space-time is considered.
\par
Since we are dealing with a primordial cosmological model, the matter content of the universe
is dominated by radiation. Hence, $\rho = \rho_0/a^4$.
In the computation that will be performed later, we will ignore the dark radiation, the
linear term for the matter density and
cosmological constant terms. These simplification will be justified a posteriori, since
these terms will come out to be smaller than the quadratic matter density one for
the energy scales relevant for the nucleosynthesis analysis.
Hence, our fundamental equation to be analysed is
\begin{equation}
\label{fe}
\biggr(\frac{\dot a}{a}\biggl)^2 = \frac{\rho^2}{M^6} = N^2\frac{\pi^4}{900}\frac{T^8}{M^6} \quad ,
\end{equation}
where we have used the fact that
\begin{equation}
\rho = N\frac{\pi^2}{30}T^4 \quad ,
\end{equation}
$N$ being the effective number of degrees of freedom of the
massless particles. Since, at the moment of the nucleosynthesis
there are photons, electron-positron pairs and three types of
neutrinos-antineutrinos, $N = 43/4$. The gravitational coupling
parameter has been redefined as $\kappa^4/36 = 1/M^6$. The
parameter $M$ defines the fundamental mass scale of brane
cosmology.
\par
The use of the equation (\ref{fe}) distinguishes strongly the
present approach from previous ones. In references
\cite{ichiki,bratt}, it has been used instead of (\ref{fe}), the
equation
\begin{equation}
\biggr(\frac{\dot a}{a}\biggl)^2 = \frac{\kappa^4}{36}\rho^2 + \frac{8\pi G}{3}\rho +
\frac{\cal C}{a^4} \quad .
\end{equation}
The main conclusion was that a large contribution of dark
radiation is allowed if ${\cal C} < 0$. But, there is one point
which must be commented. In those references, the authors have
kept the nucleosynthesis energy scales between $1\,MeV$ and
$0.1\,MeV$. They established the observational limits on the dark
radiation term and on the quadratic term by requiring that these
energy scales are not substantially affected by the presence of
these new terms. In this way, the prediction for helium abundance
(as well as for deuterium and lithium) of the standard model,
which agrees remarkably well with the observations, is not
affected.
\par
Here, the emphasis will be different. We will use (\ref{fe}) and
will leave the dynamics of the Universe dictates the new range of
energy where the nucleosynthesis takes place. In this way,
observational limits will be established for the term $M$ in
equation (\ref{fe}).
\par
The main reason for using (\ref{fe}) comes from the following
fact. The nucleosynthesis of helium depends crucially on the
reactions converting free neutrons into proton and on the decay of
free neutrons. The last effect continues until the energy of the
Universe drops much below the deuterium bindind energy,
$\epsilon_D = 2.23\,MeV$. Typically the decay of free neutrons
continue until about $T \sim 0.1\,MeV$, since the large ratio of
photons to baryons assure that even in this energy scale there are
enough photons able to dissociate the formed deuterium. Below this
energy, deuterium is not dissociated anymore and they can combine
to form the helium. This fact depends only on the statistic of the
deuterium immersed in a photonic gas, depending very little on the
dynamics of the Universe. However, the reactions converting
neutrons to protons are effective until the moment the reaction
rate is equal to the rate of the expansion of the Universe, when
the number of free neutrons due only to this effect is frozen.
\par
The equality between the reaction rate ${\cal R}$ and the rate of
the expansion
of the Universe implies
\begin{equation}
{\cal R} \sim \frac{\dot a}{a} \quad .
\end{equation}
Since, in the model dictated by equation (\ref{fe}),
\begin{equation}
\frac{\dot a}{a} \sim \frac{T^4}{\sqrt{M^6}} \quad ,
\end{equation}
and the reaction rate for the conversion of neutrons into protons is given by
${\cal R} = G_F^2T^5$, where $G_F \sim 10^{-5}\,GeV^{-2}$ is the Fermi constant of
the weak interactions,
we find that the frozen temperature is given by
\begin{equation}
\label{frozen}
T_F \sim \frac{1}{G_F^2\sqrt{M^6}} \quad ,
\end{equation}
in opposition with the standard case where
\begin{equation}
T_F^3 \sim \frac{1}{G_F^2M_{Pl}} \quad .
\end{equation}
We will treat $M$ as a free parameter, used to fit the observed
helium abundance. The final results will justify the fact that the
quadratic term in fact dominates the right hand side of the
Einstein equations. But, before computing this, let us first
describe the semi-analytical method to be used.

\section{The semi-analytical computation of the primordial nucleosynthesis}

During the thermal history of the Universe, a quark-hadron
transition occurs at a temperature higher than $T \sim 100 \,
MeV$. The primordial nucleosynthesis begins after this phase
transition. Initially, the quantity of neutrons and protons is
equal. As the Universe expands, the temperature drops, and
reactions involving the neutrons, protons, neutrinos and electrons
convert neutrons into protons. At same time, since the neutrons
are free, and unstable, they also decay into protons. The
reactions converting neutrons into protons ends when its rate is
equal to the rate of the expansion of the Universe. However, the
decay of free neutrons continues until the energy is low enough in
order the neutrons to be captured forming deuterium, from which
the helium is formed. The quantity of helium produced in this
process depends essentially on the quantity of neutrons that has
survived up to the moment they are captured to form deuterium. The
detailed analysis of all these process is quite involved,
requiring the use of numerical codes to evaluate the different
transmutation process. However, semi-analytical expressions can be
worked out if the Fermi-Dirac statistic is neglected. This has
been done in reference \cite{bernstein}, leading to values for the
helium abundance with an error of some percents compared with the
precise numerical calculation. Precision is slightly lost by using
this semi-analytical analysis, but on the other hand, the physical
meaning of the computation becomes more transparent.
\par
We will summarize the main steps and the relevant quantities. For
details, the reader is invited to address himself to the reference
\cite{bernstein}. We will consider first the standard cosmological
model, deep in the radiative phase. The ratio of neutrons with
respect to the total baryon number as function of the temperature
is
\begin{equation}
X(T) = \frac{n_n(T)}{n_n(T) + n_p(T)}
\end{equation}
where $n_n(T)$ and $n_p(T)$ are the numbers of neutrons and
protons, respectively. The main equation controlling this quantity
is
\begin{equation}
\frac{dX(t)}{dt} = \lambda_{pn}(t)(1 - X(t)) - \lambda_{np}(t)X(t) \quad ,
\end{equation}
where $\lambda_{pn}$ and $\lambda_{np}$ are the rates of conversion of protons into
neutrons and neutrons into protons respectively. The main process concerned are
\begin{equation}
\label{trans}
\lambda_{np} = \lambda(\nu + n \rightarrow p + e^-) +
\lambda(e^+ + n \rightarrow p + \bar\nu) + \lambda(n \rightarrow p + \bar\nu + e^-)
\quad .
\end{equation}
As an example, the first one is given by
\begin{equation}
\lambda(\nu + n \rightarrow p + e^-) = A\int_0^\infty dp_\nu p_\nu^2p_eE_e(1 - f_e)f_\nu
\quad ,
\end{equation}
where $A$ is a coupling constant, the $p$'s denote the momenta of
each particle involved in the process, $E$ the energy and the
$f$'s represent the Fermi-Dirac statistic factor. The last process
in equation (\ref{trans}) represents the neutron decay and it is
not considered in a first evaluation. Later, the final results
will be corrected taking it into account.
\par
In reference \cite{bernstein} the evaluation of the first two
rates in equation (\ref{trans}) is simplified by approaching the
Fermi-Dirac statistics by the Maxwell-Boltzmann one. This is
justified by the fact the temperatures concerned at the moment
these process take place are smaller than the energies of the
particles. After a lengthy evaluation of all process, we end up
with the following expression for the neutron abundance factor:
\begin{equation}
\label{frac}
X(y) = X_{eq}(y) + \int_0^ydy'e^{y'}X_{eq}^2(y')\exp[K(y) - K(y')]
\end{equation}
with the following definitions:
\begin{eqnarray}
X_{eq}(y) &=& \frac{1}{1 + e^y} \quad , \\
\label{oe1}
K(y) &=& - b\biggr[\biggr(\frac{4}{y^3} + \frac{3}{y^2} + \frac{1}{y}\biggl)
+ \biggr(\frac{4}{y^3} + \frac{1}{y^2}\biggl)e^{-y}\biggl] \quad , \\
b &=& a\biggr[\frac{45}{4\pi^3N}\biggl]^{1/2}\frac{M_p}{\tau\Delta m^2}
\quad , \quad a = 4A\tau(\Delta m)^5 \quad, \quad y = \frac{\Delta m}{T}
\end{eqnarray}
where $\Delta m =  1.294 MeV$ is the mass difference between
neutrons and protons, $a = 255$ is a pure number and $\tau$ is the
neutron lifetime. The fraction of neutrons to baryons at the end
of all those process, $\bar X$, is obtained by making $y
\rightarrow \infty$ ($T \rightarrow 0$), $\bar X = X(y \rightarrow
\infty)$. The fact that the integration is performed until $T =
0$, is not important since equation (\ref{frac}) has yet saturated
at $y \sim 10$, that is, $T \sim 0.1\,MeV$, the temperature for
which the neutrons were already captured forming deuterium, and
the nucleosynthesis process has finished.
\par
As stated before, initially the neutron decay is neglected. To correct the
final results due to
it we must evaluate the time where the capture process
forming the deuterium occurs.
This time is obtained, in principle, by evaluating the capture process until a temperature
of the order of the deuterium binding energy $\epsilon_D \sim 2.225\,MeV$.
However, lower energies (of order of $E \sim 0.1\,MeV$) must
be considered due to the fact that the enormous number of photons implies that
deuterium dissociation continues to occur even when $T_\gamma < \epsilon_D$. In peforming
this analysis, we
must take into account the evolution of the Universe, which in this case is
reflected by the equation
\begin{equation}
\label{oe}
\frac{\dot T_\gamma}{T_\gamma} = - \biggr[\frac{8\pi\rho}{3M_P^2}\biggl]^{1/2}
\end{equation}
since during the radiative phase $a \propto 1/T_\gamma$, $T_\gamma$ being the
photon temperature, which is approximately equal to the neutrino temperature
at the relevant temperature scales considered in this computation. Moreover,
the energy density is given by
\begin{equation}
\rho = N_{eff}\frac{\pi^2}{30}T^4_\nu \quad ,
\end{equation}
where
\begin{equation}
N_{eff} = N_\nu + \biggr(\frac{11}{4}\biggl)^{4/3}N_\gamma \sim 13 \quad .
\end{equation}
\par
The equation (\ref{oe}) allows to establish a relation between
temperature and time. In doing this we can convert the expression
for the neutron decay in terms of the temperature. The final
results imply that time is related with temperature by
\begin{equation}
t = \biggr[\frac{45}{16\pi^3N_{eff}}\biggl]^{1/2}\biggr[\frac{11}{4}\biggl]^{2/3}
\frac{M_P}{T_{\gamma0}^2} + t_0 \quad .
\end{equation}
The constant $t_0$ is
\begin{equation}
t_0 = \frac{11}{6N_{eff}}\biggr[(\frac{11}{4})^{1/3} - 1\biggl]t_1
\quad , \quad t_1 = \biggr[\frac{45}{16\pi^3N_{eff}}\biggl]^{1/2}\frac{M_P}{T_{\gamma0}^2}
\quad ,
\end{equation}
where $T_{\gamma0}$ is the photon temperature when the neutron
capture is accomplished. The time of capture is obtained through
some statistic considerations, and its expression is given by
solving the equation
\begin{equation}
\label{oe2} 2.9\times10^{-6}z_c^{-17/6}e^{-1.44z_c^{1/3}+z_c} \sim
1
\end{equation}
where $z_c = \epsilon_D/T$.
The final abundance is given by
\begin{equation}
X_f = \exp{(-t_c/\tau)}\bar X \quad ,
\end{equation}
where $t_c$ is the time when the neutrons are captured into
deuterium. The final helium abundance by weight ratio is
\begin{equation}
Y_4 = 2X_f \quad .
\end{equation}
The complete computation yields $Y_4 \sim 0.247$.
A more precise
computation, using numerical code, gives $Y_4 \sim 0.241$ \cite{braginsky}.

\section{Nucleosynthesis in the brane cosmology}

\begin{figure}[t]
\begin{center}
\includegraphics[scale=0.80]{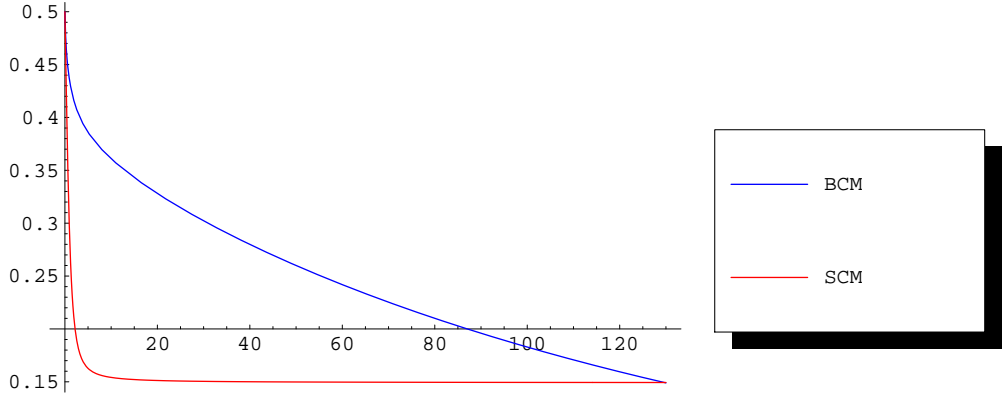}
\end{center}
\caption{The behavior of the function $K(y)$ for the standard
model and for the brane model until the capture time.}
\label{figA}
\end{figure}

All the computation of the helium abundance in a brane cosmology
definite by equation (\ref{fe}) is the same as before. The main
changes concerns the computation of the function (\ref{oe1}) and
the relation (\ref{oe2}) due to the employment of equation
(\ref{fe}) instead of the traditional Friedmann equation. Due to
the different dynamics of the Universe, driven now by equation
(\ref{fe}), the function (\ref{oe1}) is now replaced by
\begin{equation}
\label{ne1}
K'(y) = - b'\biggr\{y - \frac{12}{y}(1 + e^{-y}) - e^{-y} + 6\ln y + 6\int_y^\infty\frac{e^{-t}}{t}\,dt\biggl\} \quad ,
\end{equation}
where $b' = a\frac{M^3}{\tau\Delta m^4}\frac{30}{N\pi^2}$, all
other quantities remaining the same. An important feature of
expression (\ref{ne1}) is that it does not saturate as in the
standard case, due to the presence of the linear and logarithmic
terms. Hence, the integration must be performed until the capture
time of neutrons into deuterium, and not until $T = 0$. The
behavior of the function (\ref{ne1}), compared with the
corresponding function for the standard model, is displayed in
figure (\ref{figA}).

\par
Note that the transmutation of neutrons and protons in the brane
cosmology remains effective longer after the frozen temperature
since the ratio between the expansion of the Universe and the
reaction rate of weak interactions scales, in the brane model,
with $T$, while in the standard model this ratio scales as $T^3$.
\par
On the other hand, the capture time is also modified by the
dynamics of the Universe, since relation (\ref{oe}) is no longer
valid, being replaced by
\begin{equation}
\label{ne2}
\frac{\dot T_\gamma}{T_\gamma} = - \frac{N\pi^2}{30}\frac{T^4}{M^3} \quad .
\end{equation}
This leads to the following new expression for the capture time:
\begin{equation}
e^{z - \xi z^{1/3}}z^{-5/6} \approx 1 \quad ,
\end{equation}
where
\begin{equation}
\xi = 3\pi\alpha\biggr(\frac{m_p}{2\pi\alpha\epsilon_D}\biggl)^{1/3}
\end{equation}
with $\alpha$ being the fine structure constant and $m_p$ the
proton mass.
\par
All the computations depend on the value of the parameter $M$. The
results are displayed in table $1$.
\newline
\vspace{0.5cm}
\begin{center}
\begin{tabular}{|c|c|c|}
\hline
$1/M^3(MeV^{-3})$&X&Y\\ \hline
$10^{20}$&0.5&1\\ \hline
$10^{5}$&0.5&1 \\ \hline
1&0.5&1 \\ \hline
$10^{-5}$&0.5&1 \\ \hline
$10^{-12}$&0.5&1 \\ \hline
$10^{-13}$&0.5&0.999999 \\ \hline
$10^{-14}$&0.499999&0.999996 \\ \hline
$7.90\times10^{-21}$&0.233139&0.250547 \\ \hline
$7.38\times10^{-21}$&0.223207&0.230567 \\ \hline
$10^{-40}$& - & - \\ \hline
\end{tabular}
\end{center}
\begin{center}
{Table 1: Helium abundance in terms of the mass scale value.}
\end{center}
\par
From the table $1$, it is possible to verify that until a value
$M^6 \stackrel{<}{\sim} 10^{24}\,MeV^6$ the value of the five
dimensional coupling parameter is so high that it drives a so fast
expansion that no transmutation occurs. Hence, the nucleosynthesis
period finishes with the same quantity of neutrons it begins, that
is, $n_p = n_n$. All neutrons and protons are used to produce
helium, what explains the result $Y = 1$. On the other hand, for
values of $M^6 \stackrel{>}{\sim} 10^{80}$, the expansion is so
slow, that no neutrons are left: The number of helium produced is
zero. The values for the five dimensional coupling parameter such
that the observational limits for the helium abundance are
satisfied imply $5.0\times10^3\,GeV < M < 5.1\times10^3\,GeV$,
giving a very good agreement with the observed value, which is
$0.23 < Y^{Obs} < 0.25$ \cite{olive}.
\par
Now, we can justify the approximations fixed before. First notice
that the frozen temperature for the best values of the mass scale
is now $T_F \sim 75\,MeV$, that near the quark-hadron phase
transition temperature. This is not in principle a problem if this
phase transition occurs really a little before that energy. Let us
consider now the equation (\ref{em2}). With those best values for
$M$, and making the identification
\begin{equation}
\frac{\kappa^4}{36} = \frac{1}{M^6} \quad , \quad \frac{2\sigma}{M^6} = 8\pi G = \frac{8\pi}{M_{Pl}^2} \quad ,
\end{equation}
we have at the temperature of $T_i \sim 100\,MeV$, the following values for each term:
\begin{equation}
\frac{\rho_r^2}{M^6} \sim 6\times10^{-25}\,MeV^2 \quad , \quad
\frac{\rho_r}{M_{Pl}^2} \sim 6\times10^{-37}\,MeV^2 \quad , \quad
\Lambda \sim 10^{-49}MeV^2 \quad .
\end{equation}
Hence, under these conditions, the two last term in the right hand
side of equation (\ref{em2}) can indeed be neglected. On the other
hand, the nucleosynthesis process ends when the energy is about
$T_f \sim 0.1\,MeV$. At this energy, the terms on the right hand
side of equation (\ref{em2}) read
\begin{equation}
\frac{\rho_r^2}{M^6} \sim 6\times10^{-49}\,MeV^2 \quad , \quad
\frac{\rho_r}{M^2_{Pl}} \sim 6\times10^{-49}\,MeV^2 \quad , \quad
\Lambda \sim 10^{-49}MeV^2 \quad .
\end{equation}
This assures that the quadratic term really dominates over the linear and the constant term
during all the nucleosynthesis period, justifying a posteriori our first approximation.
Concerning the dark radiation term, we may consider that it can also be ignored during the
nucleosynthesis period since its energy density today must not exceed too much the
radiation density today. It scales like the ordinary radiation density; hence, its
value at the nucleosynthesis period must be at most of the order of the linear matter term, much
smaller than the quadratic term at least until $T \sim 0.1\,MeV$.
Notice however, that the constant term leads to a vacuum energy today of the
order $\rho_V \sim 10^{-16}\,GeV^4$, much higher than the observed value $\rho_V^{0bs} \sim
10^{-47}\,GeV^4$ \cite{carroll}. Even if this cosmological term does not spoil the nucleosynthesis
process, it is not compatible with the later evolution of the Universe. The model must
be supplemented with a mechanism to circumvent this problem through a dynamical process or
a fine tuning.

\section{Conclusions}

In this paper we analyzed the primordial production of helium
using a semi-analytical approach in the context of brane
cosmology. The main point in this computation was to consider the
predominant role of the quadratic matter density in brane
cosmology, since this term dominates at very high energy scales.
In doing so, we ignored the contribution of dark radiation. This
assumption was justified a posteriori, since dark radiation and
the ordinary linear radiation term are really a subdominant
component to energy scales down to $T \sim 0.1\,MeV$, if the
fundamental mass scale is such that $M \sim 5\times10^3\,GeV$.
\par
The main difference with respect to previous work on the subject
concerns the computation method and the energy scales where the
primordial nucleosynthesis occurs. In what concerns the
computation method, we have opted for a semi-analytical method
developed in reference \cite{bernstein}, which slightly sacrifies
the precision but furnishes a more physical insight. Concerning
the energy scales for the nucleosynthesis, we allow them to vary.
In references \cite{ichiki,bratt}, the authors kept the energy
scales as the same as in the standard cosmological model, that is
$1\,MeV < T < 0.1\,MeV$. Then, they constrained the dark radiation
term and the quadratic matter density term such that the predicted
helium abundance remains inside the observational limit. Here, on
the other hand, the energy scales for the primordial
nucleosynthesis were kept free, and we determine them, and at same
time the fundamental mass scale, in such a way that the
observational data can be reproduced.
\par
The main results can be resumed as follows. If the mass scale is to small, there is no
time for the transmutation of neutrons into protons and all baryons finish to form
helium; no or almost no hydrogen is left. This, of course, contradicts observations.
However, if the mass scale is too high, essentially all neutrons are transmuted
into protons and no helium is produced, contradicting again the observations. There is
an optimal value for the mass scale, where the observational abundance of helium can
be produced. For this optimal value, the frozen temperature for the transmutation of
neutrons into protons is of the order of $T_F \sim 75\,MeV$, while the neutrons capture time
inside deuterium remains essentially the same as in the standard scenario,
i.e, $T \sim 0.1\,MeV$. The value for the mass scale is $M \sim 5\times10^{3}\,GeV$.
\par
These results are quite reasonable. With these values, the initial
hypothesis that the contribution of dark radiation, the linear
matter density term and of the cosmological constant are
negligible are perfectly justified. However, there are two main
difficulties arising from this analysis. First, the frozen
temperature is dangerously near the typical temperature for the
quark-hadron phase transition, which is not modified by the change
in the dynamics exactly because it is a phase transition. But,
there are yet many discussions on the exact value of the
temperature at which this phase transition occurs \cite{dominik}.
Second, the value of the cosmological constant, even if
unimportant for the nucleosynthesis process, is too high in view
of current observational limits on the vacuum energy. Some fine
tuning is necessary in order to avoid problems in this sense. But
this seems to be a common feature of many brane models. It is
important to stress, however, that the analysis performed here is
somehow a simplification of the problem: A more complete analysis
should taken into account the contribution of all terms,
constraining the mass scale in presence of the linear density term
and dark energy. But this approach asks for important
modifications of the semi-analytical method employed here. In any
case, we have shown that a scenario where the quadratic term
dominates during nucleosynthesis is possible, provided that the
relevant energy scales are modified.
\par
In our point of view, the main contribution of this analysis
concerns the value predicted for the mass scale. It is somehow
below other limits. In reference \cite{ichiki} for example, the
authors have determined a bound such that $M > 13\,TeV$. Our
results are marginally consistent with this bound and, moreover,
they set the fundamental mass scale of branes near the terrestrial
accelerators energy limit, and very near the scales of other
fundamental interactions. If we had taken into account the other
contributions, we could obtain a higher value for the mass scale.
In view of a more general analysis, the result found here for the
mass scale may be seen as lower bound.
\newline
 \vspace{0.5cm}
 \newline
 {\bf Acknowledgments:} We thank Patrick Peter, Roberto Colistete Jr.
 and S\'ergio V.B. Gon\c{c}alves for their suggestions and
 comments on this text. We thank CNPq (Brazil) for partial financial support.
 J.A.O. Marinho has received a fellowship from PIBIC/UFES/CNPq
 during the elaboration of this work.

\end{document}